\pdfoutput=1
\documentclass[aps,pre,nofootinbib,twocolumn]
{revtex4}
\usepackage{graphicx}
\usepackage{amssymb}
\usepackage{amsmath}
\usepackage{caption}
\usepackage{multirow}
\usepackage{array}
\usepackage{float}
\usepackage{subfigure}
\usepackage{color}
\usepackage{adjustbox}
\usepackage{lipsum}

\begin{document}


\title {Effect of Channel Noise in  Synchronization and Metabolic Energy Consumption in Unidirectionally Coupled Neurons: Drug Blocking  of Sodium and Potassium Channels}

\author{Krishnendu Pal$^{1,2,+}$ and Gautam Gangopadhyay$^{2,}$}
\email{+pckp@iacs.res.in,*gautam@bose.res.in}

\affiliation{$^{1}$Indian Association for the Cultivation of Science, Jadavpur, Kolkata-700032, India.\\$^2$S N Bose National Centre for Basic Sciences, Block-JD, Sector-III, Salt Lake, Kolkata-700106, India.}


\date{\today, to be submitted soon}

\begin{abstract}
In this work the  stochastic generalization of single  Hodgkin-Huxley neuron  is further extended to unidirectionally coupled neurons. Our main focus is to elucidate  the role of channel noise in the kinetics and  energetics of spiking of action potential and the synchronization  between two coupled neurons.  We have found that the size of the patch is playing the pivotal role in synchronization and metabolic energy consumption. For example, there exists three different patch size ranges in which coupled neuron system behaves in a different manner from noise enhanced phase to dead range state before reaching the deterministic limit. We have also found that the sodium and potassium channel blockers have characteristic kinetic and energetic effects on synchronization process and metabolic energy consumption rate which has been validated with the simulated data. 
\end{abstract}


\maketitle

\section{Introduction}
\label{s1}
The generation and conduction of action potentials is the fundamental means of cellular communication in the nervous system. Generation of action potential to maintain neural activity requires high  metabolic energy\cite{Clarke, Attwell, Siekevitz}. In humans, the brain has only 2\% of the body mass but it consumes 20\% of the human metabolic energy\cite{Clarke} which is a large fraction. The generation of the firing sequences of action potential to transmit information\cite{Attwell} consumes a large fraction of brain's total metabolic energy consumption.  The action potential involves influx of Na$^+$ ions and efflux of K$^+$ ions through voltage-gated ion channels, which charge the membrane capacitance to the peak of the action potential and then discharge it back to resting potential; thereby the whole process consumes huge energy.  Ion pumps remove ions through the cell membrane against their concentration gradient consuming ATP molecules. There are three basic reasons why action potential demands huge energy. Firstly, making a robust signal requires the membrane capacitance  to be charged by more than 50 mV to the peak of the action potential. Secondly, the total area of the membrane covered by action potential during its propagation through axons, collaterals and dendrites is very large and so the capacitance must be charged to the peak voltage. Thirdly, the flux of Na$^+$ and K$^+$ ions exceeds the minimum required to charge the membrane to peak potential because the Na$^+$ and K$^+$ currents overlap\cite{Sengupta, Crotty}. Energy consumed by a neuron during an action potential is estimated by recording some action potentials and then computing theoretically  the number of sodium ions required to enter into the cell to produce the same membrane depolarization\cite{Sengupta, Crotty, Alle, Hod}. The Hodgkin-Huxley\cite{hodg1} circuit model is the most used model for the study of neurons 
 using which the average metabolic energy consumption can also be estimated by calculating the total derivative of the electrochemical energy in the circuit\cite{Torrealdea, moujahid1}.


However random disturbances of signals, termed as noise create fundamental problems for information processing, synchronization and affect all functioning aspects of nervous-system. The noise sources can be of various types, such as sensory noise, cellular noise, electrical noise, synaptic noise, motor noise\cite{faisal}, extrinsic photon shot noise etc. Noise is also generated intrinsically by the stochastic opening and closing of voltage gated ion channels found in neural membranes\cite{faisal, Senguputa1, neher, lecar, Lecarand}. Stochasticity in voltage-gated ion channels is caused by random fluctuations between different conformational states due to thermal agitation\cite{white}. Stochastic models have shown that channel noise can account for variability in the action potential threshold at nodes of Ranvier\cite{rubinstein} and the reliability of action potential initiation in membrane patches\cite{Strassberg, walloe, chow}. Patch-clamp experiments in vitro show that channel noise in the dendrites and in the soma produces membrane-potential fluctuations that are large enough to affect action potential timing\cite{Jacobson, Kole}. At the site of action potential initiation i.e. at the soma or at the axon hillock channel noise  literally  affect the timing of action potentials although there exist comparatively large number of ion channels present at these sites\cite{rubinstein}. Stochastic simulations show that the smaller number of ion channels that are open at the action potential threshold actually determines the timing precision, not the number of ion channels that are open at the peak of the action potential\cite{van}. The effect of channel noise increases dramatically as neurons become smaller\cite{faisal1}. The membrane potential is affected by the opening of ion channels in proportion to the membrane's input resistance, which increases rapidly with decreasing diameter of the neuron\cite{rall}. The input resistance of axons of diameter  bellow 0.3 $\mu m$  gets large enough, that at the resting potential spontaneous opening of single Na+ channels produce `Na+ sparks' which can trigger action potentials even in the absence of any other stimulus. These `rogue’ action potentials become exponentially more frequent as  diameter of the axon decreases.  Axons below 0.08-0.10 $\mu m$ diameter are useless for communication. This lower limit matches with the smallest diameters of axons across species\cite{faisal1}. Channel noise also affects action potential  propagation in axons, producing trial-to-trial variability in action potential timing\cite{faisal2}.  

The nature, amount and impact of channel noise in the nervous system have been addressed since the 1970s in many experimental and computational methods both quantitative and qualitatively.  But the effect of patch size or channel noise in the energetics of action potential in single neuron and in coupled neurons has been grossly overlooked.  To explore the effect of patch size or channel noise in metabolic energy consumption and in the synchronization  between two unidirectionally coupled neurons, we have considered here a simple coupled Hodgkin-Huxley neuron system connected via gap junction channels of particular conductance. We first calculate the metabolic energy consumed by the ion channels using single Hodgkin-Huxley circuit model. Then we incorporated the channel noise in the unidirectionally coupled neurons and studied the kinetics and energetics of the spiking activity. For simplicity we have considered both the neurons have similar density of ion channels and they are of same size. Next the effect of sodium and potassium channel blockers have been extensively investigated as a part of the patch size study.   To be more specific in this work we have asked the following questions: 
(1) How does the internal channel noise or the size of a neuron contribute to the metabolic energy consumption in both single and coupled neuron system? 
(2) How does the channel noise kinetically and energetically affect the neural synchronization process? 
(3) How does the  sodium blockers, potassium blockers and total blockers act on the action potential response and metabolic energy consumption. 
(4) What are the qualitative and quantitative differences between these three types of drug blockers?    


The layout of the paper is as follows. In section (\ref{s2}) we have discussed the deterministic description of metabolic consumption of a single neuron in presence of external current. Then in section (\ref{s3}) we have discussed the kinetic scheme and energetics of the unidirectional coupling. In section (\ref{s4}) the channel noise is introduced in the coupled neuron and in various subsections we have studied the effect of patch size on synchronization and average metabolic energy consumption. In section (\ref{s5}) we have explored the effect of three types of channel blockers on metabolic energy consumption and validated with the simulated result. Finally the paper is concluded in section (\ref{s6}).


\section{Single Hodgkin-Huxley Neuron: Deterministic description}
\label{s2}

We begin with the well known circuit model of Hodgkin-Huxley equation\cite{hodg1} for the  action potential in squid giant axon,
\begin{widetext}
\begin{equation}
C_m \frac{d}{dt} V(t)+ G_K(t)(V(t)-E_K)+G_{Na}(t)(V(t)-E_{Na})+G_{L}(V(t)-E_L)=I_{ext}(t),
\label{hh}
\end{equation}
\end{widetext}
where V(t) is the membrane potential. Parameters are taken from the papers of Schmid,  Goychuk and  Hanggi\cite{hangii1, hangii2, hangii3} and their descriptions are given in Table \ref{t1}. Conductance $G_K(t)$ and $G_{Na}(t)$ are given as follows,
\begin{equation}
G_K(t)= g^{max}_K n^4\hspace{0.1cm} \text{and} \hspace{0.1cm}G_{Na}(t)= g^{max}_{Na} m^3h,
\label{cond}
\end{equation}
where n, m and h are the well know gating variables of potassium and sodium channels which describe the mean ratios of the open gates of the working channels. The factor $n^4$ and $m^3h$ are the mean portions of the open ion channels within the membrane patch.The dynamics of the opening probabilities for the gates are given by
\begin{equation}
\dot{x}= \alpha_x(V)(1-x)-\beta_x(V)x, \hspace{0.2 cm} x=n,m,h.
\label{gate}
\end{equation}
The expressions of the voltage dependent rates\cite{hangii1, hangii2, hangii3} are given as follows $\alpha_m(V)=(0.1(V+40))(1-\exp[-(V+40)/10])^{-1}$, $\beta_m(V)= 4\exp[-(V+65)/18]$, $ \alpha_h(V)=0.07 \exp[-V+65)/20]$, $\beta_h(V)={1+\exp[-(V=35)/10]}^{-1}$, $ \alpha_n(V)=(0.01(V+55))(1-\exp[-(V+55)/10])^{-1}$, $\beta_n(V)=0.125\exp[-(V+65)/80]$.

\begin{center}
\begin{table}[h]
\caption{Parameters of Hodgkin-Huxley equation\cite{hangii1}.} 
\begin{adjustbox}{max width=0.48\textwidth}
\begin{tabular}{|c|c|c|}
\hline

\hspace{0.5cm}$C_m$\hspace{0.5cm}  & \hspace{0.5cm} Membrane capacitance\hspace{0.5cm} & \hspace{0.5cm}1 $\mu$F/cm$^2$\hspace{0.5cm}  \\

\hspace{0.5cm}$E_K$\hspace{0.5cm}  & \hspace{0.5cm} K$^+$  reversal potential\hspace{0.5cm} & \hspace{0.5cm}-77.0 mV\hspace{0.5cm}  \\

\hspace{0.5cm}$\rho_{K}$\hspace{0.5cm}  & \hspace{0.5cm} K$^+$ channel density\hspace{0.5cm} & \hspace{0.5cm}18 channels/ $\mu$m$^2$\hspace{0.5cm}  \\

\hspace{0.5cm}$g_K^{max}$\hspace{0.5cm}  & \hspace{0.5cm} Maximal K$^+$ channel conductance\hspace{0.5cm} & \hspace{0.5cm}36.0 mS/cm$^2$\hspace{0.5cm}  \\

\hspace{0.5cm}$\gamma_K$\hspace{0.5cm}  & \hspace{0.5cm} Single K$^+$  channel conductance\hspace{0.5cm} & \hspace{0.5cm}20 pS\hspace{0.5cm}  \\
\hline

\hspace{0.5cm}$E_{Na}$\hspace{0.5cm}  & \hspace{0.5cm} Na$^+$  reversal potential\hspace{0.5cm} & \hspace{0.5cm}50.0 mV\hspace{0.5cm}  \\

\hspace{0.5cm}$\rho_{Na}$\hspace{0.5cm}  & \hspace{0.5cm} Na$^+$  channel density\hspace{0.5cm} & \hspace{0.5cm}60 channels/$\mu$m$^2$ \hspace{0.5cm}  \\

\hspace{0.5cm}$g_{Na}^{max}$\hspace{0.5cm}  & \hspace{0.5cm} Maximal Na$^+$  channel conductance\hspace{0.5cm} & \hspace{0.5cm}120.0 mS/cm$^2$\hspace{0.5cm}  \\

\hspace{0.5cm}$\gamma_{Na}$\hspace{0.5cm}  & \hspace{0.5cm} Single Na$^+$  channel conductance\hspace{0.5cm} & \hspace{0.5cm}20 pS\hspace{0.5cm}  \\

\hline
\hspace{0.5cm}$E_L$\hspace{0.5cm}  & \hspace{0.5cm} Leak reversal potential\hspace{0.5cm} & \hspace{0.5cm}-54.4 mV\hspace{0.5cm}  \\

\hspace{0.5cm}$g_L$\hspace{0.5cm}  & \hspace{0.5cm} Leak conductance\hspace{0.5cm} & \hspace{0.5cm}0.3 mS/cm$^2$\hspace{0.5cm}  \\

\hline

\end{tabular}
\label{t1}
\end{adjustbox}
\end{table}
\end{center}

The total electrical energy accumulated in
the circuit at a given moment in time is\cite{moujahid1}
\begin{equation}
H(t)=\frac{1}{2} C_m V^2+H_{Na}+H_{K}+H_{L},
\end{equation}
where $\frac{1}{2} C_m V^2$ is the  electrical energy  accumulated by the capacitor and the last three terms represent the energies in sodium, potassium and leak  batteries, respectively. The electrochemical energy accumulated in the batteries are potentially unlimited as the exhaustion of the batteries are not considered here. In the real neuron, the nutrients consumed with food actually prevents the ion pumps from getting exhausted. 
The rate at which the batteries supply electrical energy to the circuit is equal to the electromotive force of the batteries multiplied by the electrical current through the batteries. The total derivative with respect to time of the above energy is given by\cite{moujahid1}
\begin{equation}
\dot{H}(t)=CV\dot{V}+ I_{Na}E_{Na}+I_KE_K+I_LE_L,
\label{en1}
\end{equation}
where 
$$I_{Na}=g^{max}_{Na} m^3h(V-E_{Na}),$$ 
$$ I_{K}=g^{max}_{K} n^4(V-E_{K}),$$ and 
\begin{equation}
I_{L}=g_{L} (V-E_{L})
\label{cur}
\end{equation} 
are the sodium, potassium and leakage currents. Substituting equation (\ref{hh}) into (\ref{en1}) one can obtain the following relation,
\begin{equation}
\dot{H}=VI_{ext}-I_{Na}(V-E_{Na})+I_K(V-E_K)+I_L(V-E_L).
\end{equation}

Now substituting equations (\ref{cur}) into (\ref{en1}) we get
\begin{widetext}
\begin{equation}
\dot{H}=VI_{ext}-g^{max}_{Na} m^3h(V-E_{Na})^2-g^{max}_{K} n^4(V-E_{K})^2-g_{L} (V-E_{L})^2.
\label{en2}
\end{equation}
\end{widetext}
The above equation is the total derivative of the electrochemical energy in the neuron. The
first term in the right-hand summation represents the electrical power supplied to the neuron via the different junctions reaching the neuron such as synapse and the other three terms of the summation
represent the total metabolic energy per consumed by all the three types of ion channel per second. The metabolic consumption of the Hodgkin-Huxley neuron is calculated evaluating (\ref{en2}) at different values of the external current $I_{ext}$.
\begin{figure}[h!]
\centering
\includegraphics[width=9.0 cm,keepaspectratio]{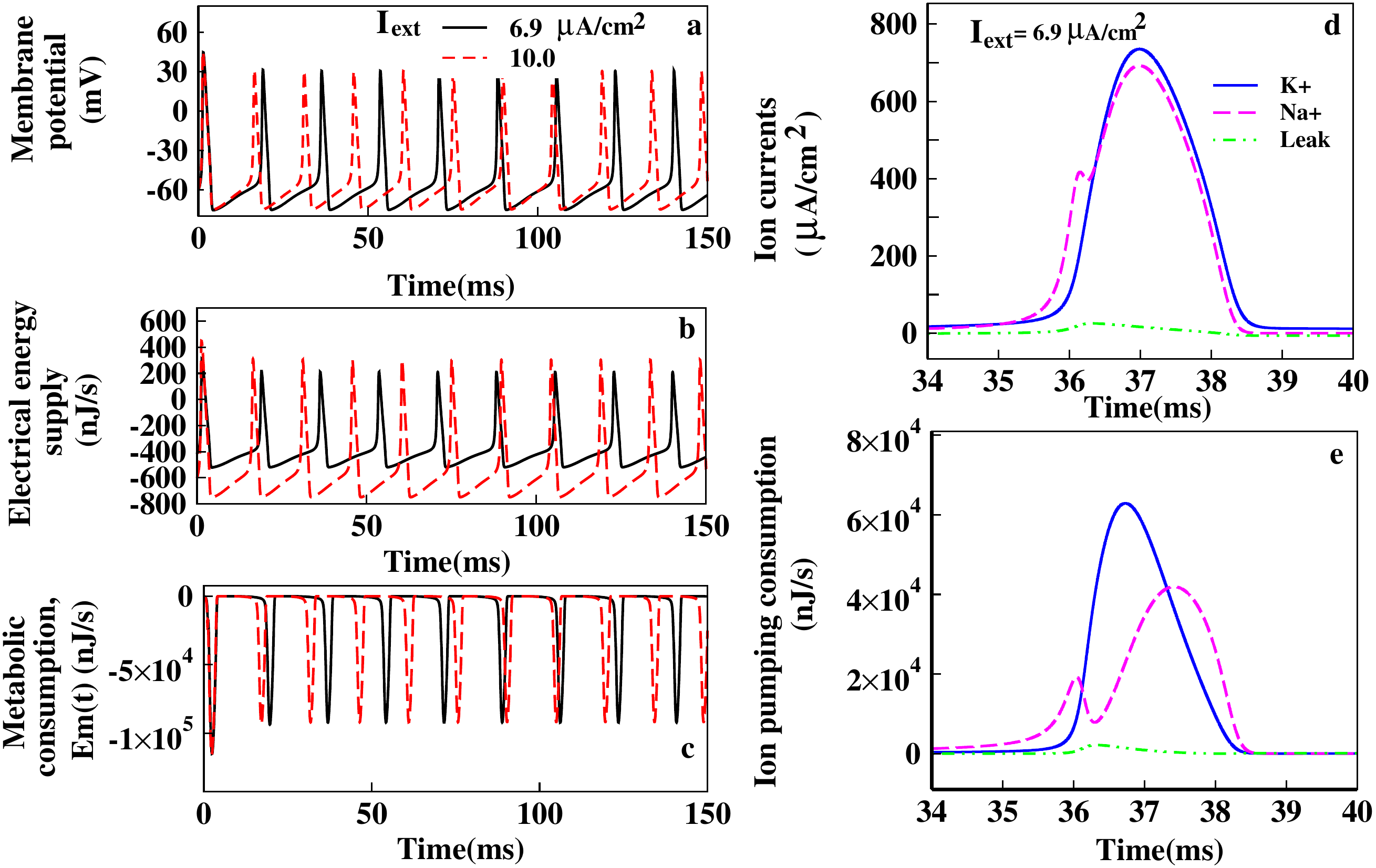}
\caption{Energy Utilization in Hodgkin-Huxley Neuron at $I_{ext}=6.9$ and $10.0 \mu A/cm^2$. In fig. (a) the action potentials are plotted. In fig. (b) energy supplied/second to the neuron is plotted. In fig. (c) the total metabolic energy  consumption rate  by all the three ion channels are plotted. (d) Currents of sodium, potassium, and
leaks for a single action potential is shown here. Sodium current is negative but
to compare with others it is plotted with  positive sign. (e) Metabolic energy consumption corresponding to all the ion currents are plotted in positive axis,  although all of them are negative.}
\label{meta}
\end{figure}

For computation we have used second order Runge-Kutta method for solving the differential equations (\ref{hh})-(\ref{gate}) . The initial voltage at which the steady states $n_{\infty}, h_{\infty}, m_{\infty}$ were calculated is taken as -70 mV. Then the stimulus voltage was provided at -60 mV at two different external current, $I_{ext}=6.9$ and $10.0$ $\mu A/cm^2$.  Then evaluating equation (\ref{en2}) we have  calculated  the electrical energy supplied to the neuron and the metabolic energy consumption, Em(t), for all the three ion channels.

In fig. \ref{meta}(a) we have plotted the  action potentials as a function of time(ms) in presence of external current, $I_{ext}=6.9$ and $10.0 \mu A/cm^2$ . With increasing $I_{ext}$ the rate of spiking activity increases. In fig. \ref{meta}(b) the rate, $V I_{ext}$ at which the electrical energy is supplied to the neuron is plotted. In fig. \ref{meta}(c) the total metabolic energy consumption rate of the neuron by its ion channels are represented. 
It is seen from the peaks of figure \ref{meta}(b) and (c) that the electrochemical energy consumption  is much greater than the energy supplied to the neuron. This is supported by the simulation result\cite{moujahid1} for which rate of energy is  replenished
by the ion pumps and metabolically supplied by hydrolysis of ATP molecules in order to maintain the neuron's activity. 

In figure \ref{meta}(d) sodium, potassium, and leak currents for a single action potential is shown. The negative sodium current\cite{kp2} is plotted on positive axis  to compare the current. The area of sodium and potassium currents are almost the same which means they neutralize each other to the extent of their mutual overlap and thus the net membrane current is much smaller. The total number of Na$^+$ or K$^+$ ions  that permeate the membrane during the action potential is proportional to the area under the  curves. In figure \ref{meta}(e)  the electrochemical energy consumption  associated with
each of the ion currents are shown. The energy consumptions  are actually negative which are plotted here in positive axis. The total metabolic consumption of the neuron in
generating one action potential is directly the sum of these
three components. 

\section{Unidirectionally Coupled Two Hodgkin-Huxley Neurons via Gap Junction: Deterministic Description}
\label{s3}

So far we have seen the energetic balance in a single neuron. Now we  focus on a system where two neurons are coupled to each other. In real system neurons are coupled to each other via electrical synapses or gap junctions and through chemical synapses. We have considered here on the electrical synapse. Electrical synapses owing to have very simple mechanism results in fast or robust signal transmission but can produce only simple behaviors, not al the complex processes where chemical synapses do well\cite{kandel}.  
 As in gap junction channels there is no need for receptors to recognize chemical messengers, signal transmissions here is more rapid than that in chemical synapses which are generally the most abundant kind of junctions between neurons. Electrical neurotransmission in electrical synapses is less modifiable than chemical neurotransmission as  they do not involve neurotransmitters. The response is always the same sign as the source and the relative speed of electrical synapses allow for many neurons to fire synchronously\cite{bennett, kandel}.

\subsection{Kinetic scheme of unidirectionally coupled Neurons}
To mathematically express the unidirctionally coupled neurons it is considered that both neurons obey two sets of  Hodgkin-Huxley equations (\ref{hh}) with an addition of a coupling term affecting the postsynaptic neuron only\cite{moujahid1}. We have considered that the two neurons are initially in states which are very close to each other. For that reason we have kept the parameters of the master neuron same as described in Table \ref{t1} and reduced the parameters of the slave neuron by 3\%, such as	$C_{2m}=0.97$, $E_{2K}=-74.69$, $E_{2Na}=48.5$, $E_{2L}=-52.768$, $g_{2L}=0.291$, $g_{2Na}^{max}=116.4$, $ g_{2K}^{max}=34.92$. With such parameters both neurons can not share time solutions or they can not be synchronous. The control action has been implemented in the slave neuron or post synaptic neuron to bring it in synchronization with the master or presynaptic neuron. It is given as the junction current $I_{Junction}=K_{sync}(V_1-V_2)$, where $K_{sync}$ is the constant conductance of the gap junction or coupling strength in $mS/cm^2$. The junction current, $K_{sync}(V_1 - V_2)$ can only flow into the slave neuron through the electrical junction from master neuron making it   unidirectional. 
The difference in electrical potential in the coupled two neuron system is maintained by an amplifier  with very large entrance impedance
that restricts the energy going back to the first neuron whereas the energy that the second neuron needs at the junction is provided by the amplifier. The slave neuron can not affect the response of the master neuron. We consider that the  membrane current of presynaptic neuron, $I_{stimulus}(t)$ is induced by a Gaussian noise, $\xi(t)$ of mean 0.0 $\mu$A
and variance 9.0 $\mu$A. Thus here $I_{stimulus}(t)=I_{ext}+\xi(t)$ and through out the rest we have taken $I_{ext}=6.9\mu A/cm^2 $.   The postsynaptic neuron is exposed to a total
current, $I_{noise}$  induced by a Gaussian noise of mean 0$\mu$A  and variance 1 $\mu$A. This noise is attributed to the mean of erratic signals coming to the postsynaptic neuron from all other synapses that we do not specifically consider here. Thus the Hodgkin-Huxley equation for the master and slave neuron stands as,
$$C_{1m} \dot{V_1}=I_{stimulus}(t)-I_{1Na}-I_{1K}-I_{1L}$$
\begin{equation}
C_{2m} \dot{V_2}=I_{noise}(t)-I_{2Na}-I_{2K}-I_{2L}+I_{Junction}.
\label{uni}
\end{equation}

\begin{figure}[h]
\centering
\includegraphics[width=8.0 cm,keepaspectratio]{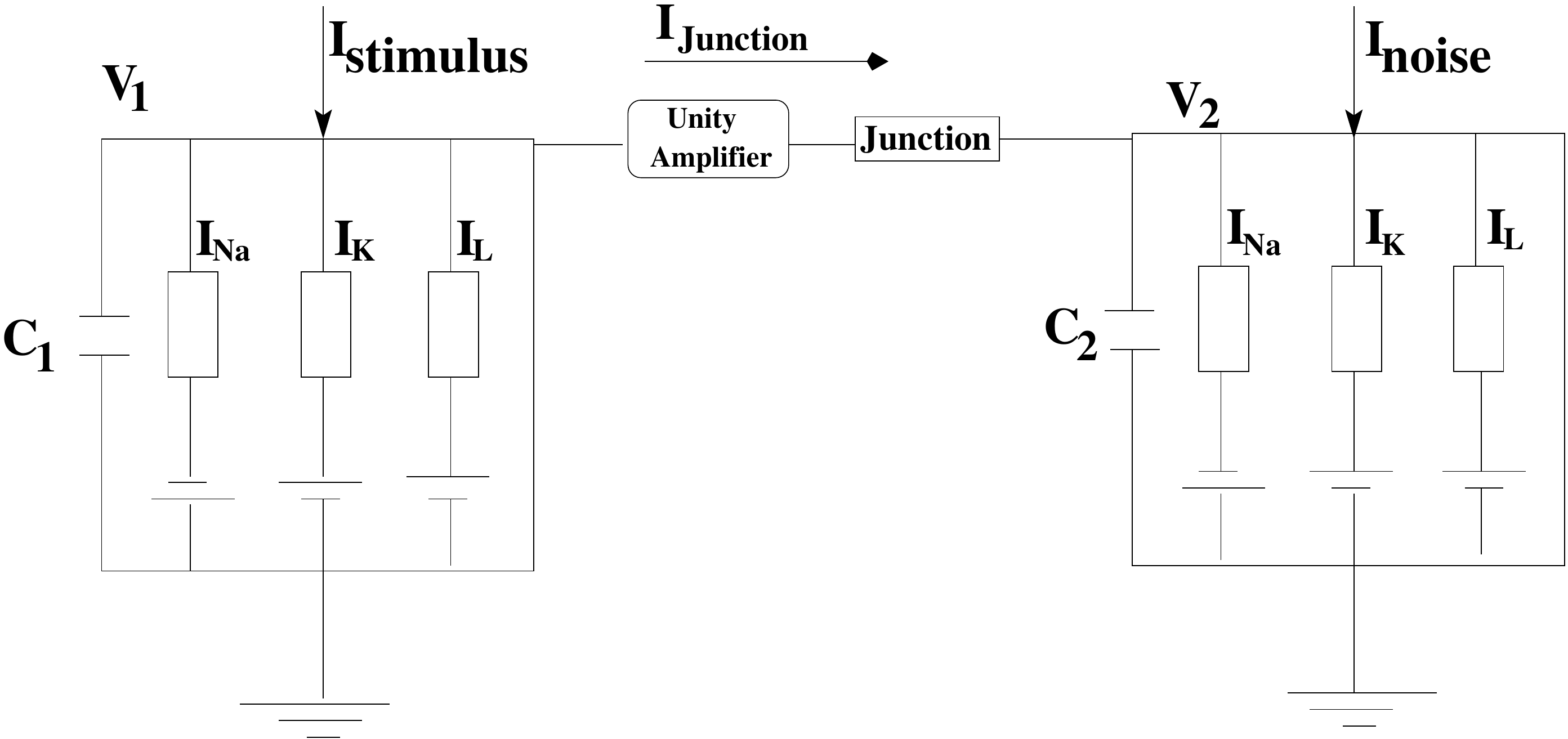}
\caption{Model of Unidirectional coupling  of two Hodgkin-Huxley neurons where the master neuron is unaffected by the second or slave neuron by the nature of the amplifier.}
\label{gap}
\end{figure}

\subsection{Energetic description of the unidirectionally coupled neurons }
As the electrical energy of the slave neuron is only affected by the coupling, it is expressed as,

\begin{equation}
H_2(t)=\frac{1}{2}C_{2m} V_2^2+ H_{2Na}+H_{2K}+H_{l}+H_{Amplifier}.
\end{equation}
The first term in  the summation is the accumulated electrical energy in the capacitor and the second, third and the fourth term upon summing up gives the total metabolic consumption of the slave neuron. The last term is the  energy available in the gap junction or the amplifier. The total energy derivative in the second neuron is given by\\
$\dot{H_2}=V_2I_{noise}-g^{max}_{2Na} m_2^3h_2(V_2-E_{2Na})^2$\\

\hspace{0.5cm}$-g^{max}_{2K} n_2^4(V_2-E_{2K})^2-g_{2L} (V_2-E_{2L})^2$\\
\begin{equation}
+ K_{sync}V_2(V_1-V_2)+ K_{sync}V_1(V_1-V_2).
\label{2en}
\end{equation}

The last two terms represent  energy balance in the junction.  Among the last two terms the first one corresponds to the energy consumed at the postsynaptic site of the junction and the second term corresponds to the energy
contributed by the amplifier. No  energy comes from the master neuron except information\cite{moujahid1}.

\section{Effect of Patch Size or Channel Number Fluctuation in Unidirectionally Coupled Neuron System}
\label{s4}

Until now we have considered only the deterministic description of the neurons. But the Hodgkin Huxley model operates on the average number of open channels disregarding the corresponding number fluctuations or channel noise. These fluctuations are inversely proportional to the number of ion channels. Thus Hodgkin Huxley model is valid only within a large system size. But in actual neuron in a finite patch size there exists a finite number of sodium, potassium and other ion channels. So the role of internal fluctuations can not be  neglected apriori . Channel noise alone can give rise to a spiking activity even in the absence of external stimulus\cite{hangii1, hangii2}.

\subsection{Stochastic generalization of Hodgkin-Huxley model: Langevin description}   

We use here Fox and Lu's\cite{fox, fox-lu} system size expansion method which uses stochastic differential equation or Langevin description to replicates the behavior of the Markov
chain model\cite{chow, walloe} with high accuracy and it is  computationally less exhaustive\cite{hangii1, hangii2, hangii3, goldwyn1}. Here the dynamics of the  gating variables are considered to be noisy as follows,
\begin{equation}
\dot{x}= \alpha_x(V)(1-x)-\beta_x(V)x+ \eta_x(t), \hspace{0.2 cm} x=n,m,h,
\label{lang1}
\end{equation}
where the terms with $\eta_x(t)$ are independent Gaussian white noise with zero mean which takes into account of the fluctuations of the number of open gates. The noise strengths depend on the membrane voltage. The noise correlations have the following form for an excitable membrane patch with $N_{Na}$ number of sodium and $N_K$ number of potassium ion channels\cite{fox, fox-lu}, respectively
\begin{equation}
<\eta_x(t)\eta_x(t')>=\frac{2}{N_i} \frac{\alpha_x\beta_x}{\alpha_x+\beta_x} \delta(t-t') \hspace{0.2cm} i= Na,K,
\label{lang2}
\end{equation}
where $N_{Na}=\rho_{Na} A$ and $N_K=\rho_K A$ are the numbers of sodium and potassium ion channels in a particular patch size of area A. 

\vspace{0.5cm}
\begin{figure}[h!]
\centering
\includegraphics[width=9.0 cm,keepaspectratio]{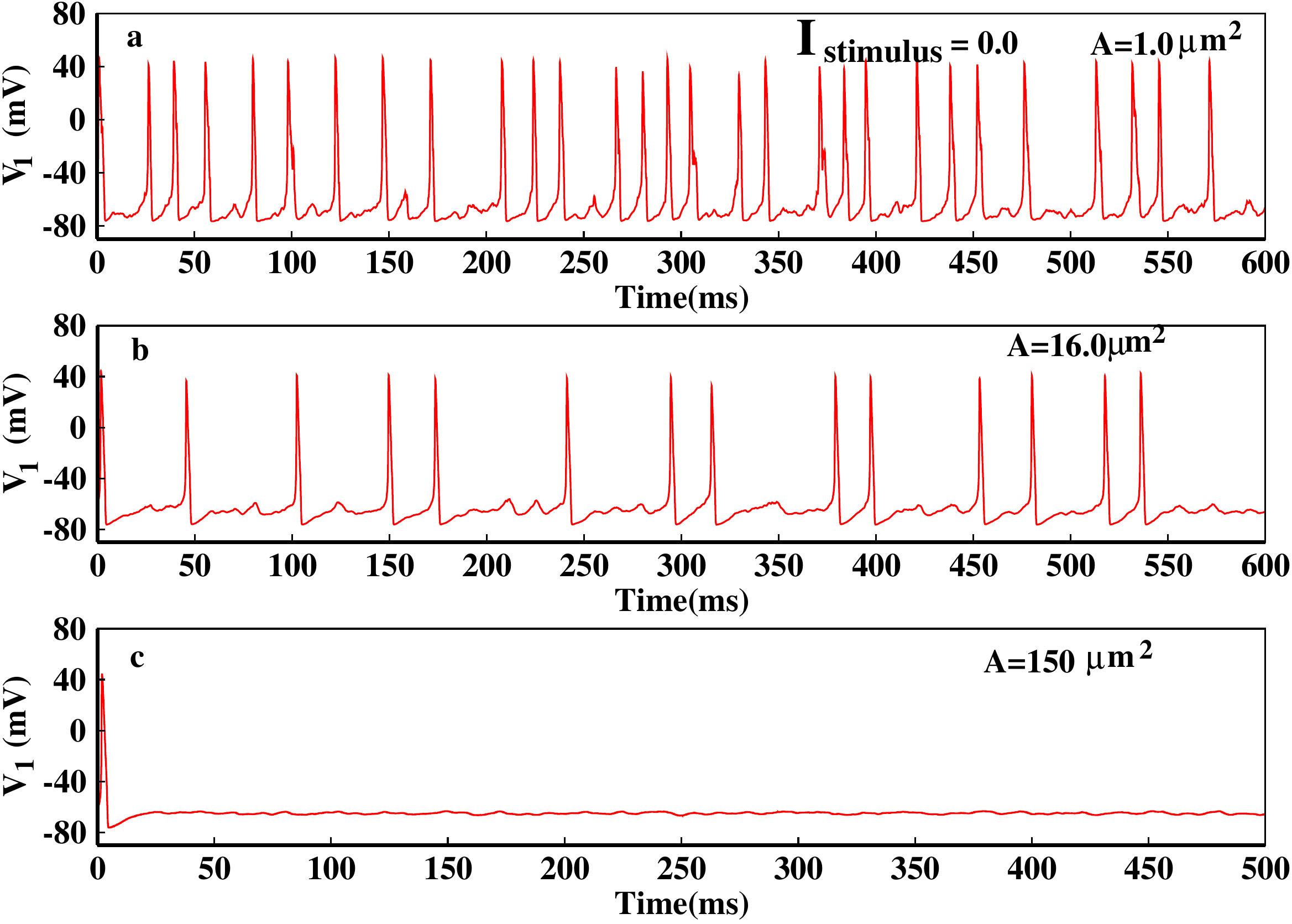}
\caption{Spontaneous generation of action potentials  using Langevin description for a single neuron with initial input of $ V_1$ =-60 mV and the steady state values of the gating variables, $x_{\infty}$ are calculated at V=-70 mV.}
\label{snl}
\end{figure} 

In figure (\ref{snl}) we have shown the effect of incorporation of the channel noise into the master neuron only, in absence of any external current, i.e. $I_{stimulus}=0.0$. Here we show that with decreasing patch size, the spiking activity increases. Thus it is seen that channel fluctuations or channel noise is sufficient to generate spontaneous action potential. With increasing path size system behaves similar to the deterministic dynamics of Hodgkin and Huxley action potential.  

Next we have incorporated the channel noise into both the master and slave neurons. For simplicity we consider both of them with equal number of sodium and  potassium channels. Here we have used  similar set of equations as in equation (\ref{lang1}) and (\ref{lang2}) for both the neurons and solved equation (\ref{uni}).

\begin{figure}[h!]
\centering
\includegraphics[width=9.0 cm,keepaspectratio]{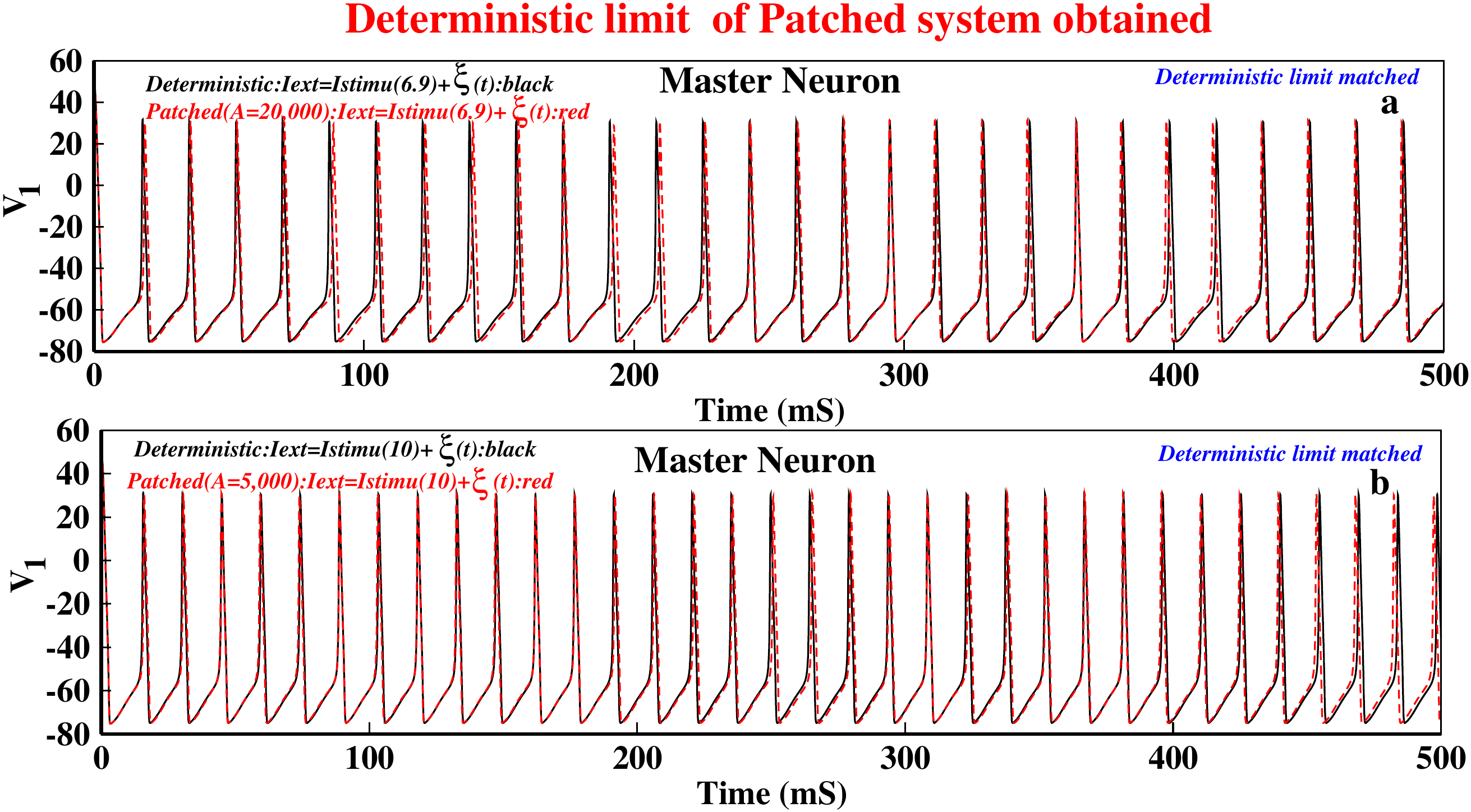}
\caption{Standardization or cross verification of the results at deterministic limit. For $I_{ext}=6.9$ $\mu A/cm^2 $ and $I_{ext}=10.0$ $\mu A/cm^2 $ the deterministic limit is achieved at A=20,000 and 5,000$\mu m^2$, respectively. } 
\label{match}
\end{figure}

To convince our result to be meaningful one can cross verify the result in the following way. We solved the deterministic Hodgkin -Huxley equation (\ref{hh}- \ref{gate}) in presence of noisy external current, $I_{stimulus}(t)=I_{ext}+\xi(t)$, as given earlier in the master neuron with similar set of Gaussian random numbers for both the deterministic and patched programs and varied the patch size from 100 to 20,000 $\mu m^2$. We have found that for an external current of $I_{ext}=6.9$ $\mu A/cm^2 $ both the deterministic and the patched solutions exactly match each other as seen from figure \ref{match}(a). It is seen that as we increase the external current the deterministic limit is achieved within a patch size of A=5,000 $\mu m^2$ as seen from figure \ref{match}(b). Here we have continued our study with $I_{ext}=6.9$ $\mu A/cm^2 $ taking A=20,000 $\mu m^2$.

\subsection{Kinetic picture of synchronization}

\begin{figure}[h!]
\centering
\includegraphics[width=9.0 cm,keepaspectratio]{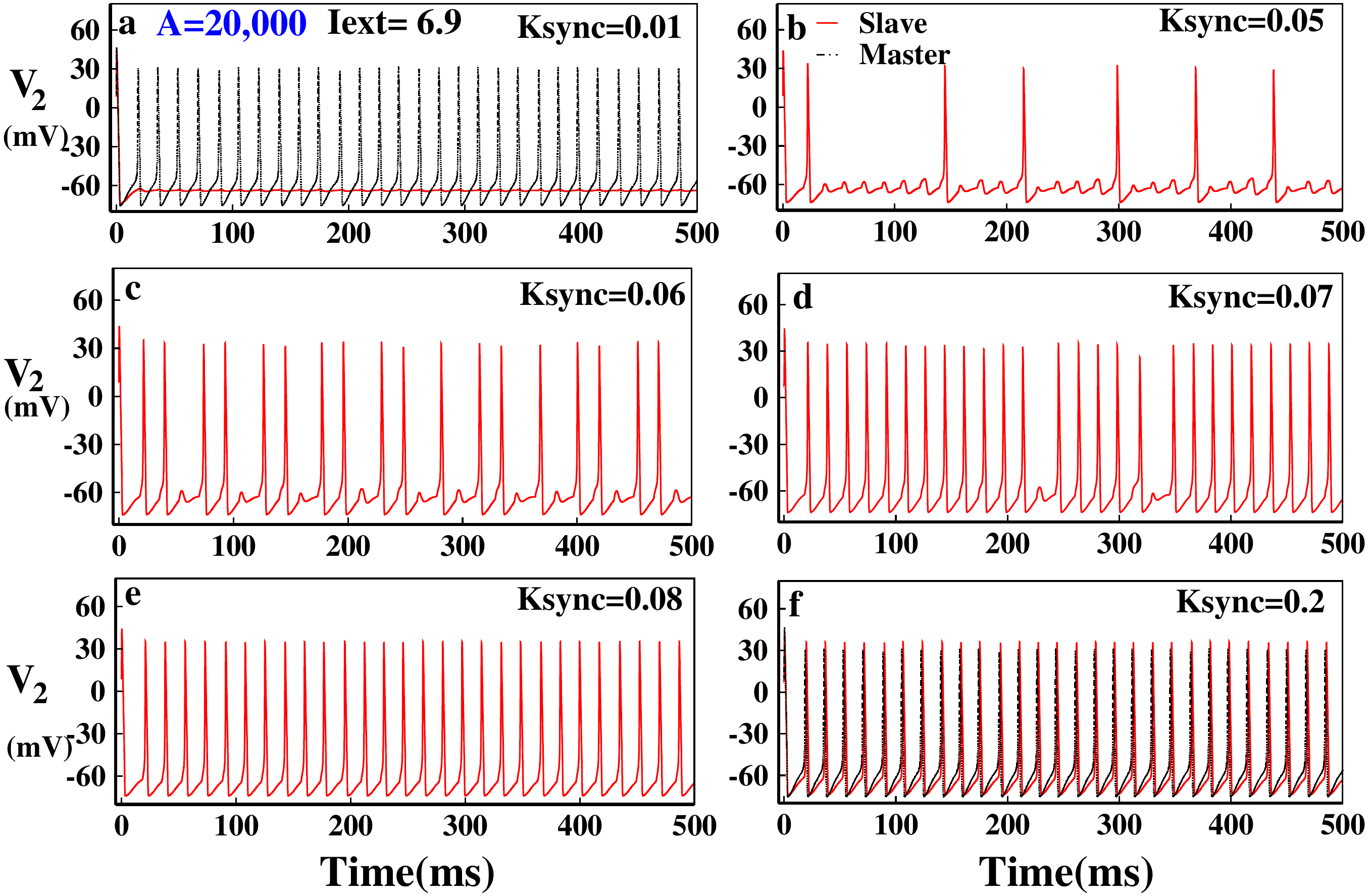}
\caption{Action potential, $V_2(t)$ trains of the slave neuron at different $K_{sync}$ values, for A=20,000$\mu m^2$, $I_{stimulus}(t)=I_{ext}(=6.9)+\xi(t)$. In figure (a) and (f) the black dashed curves which correspond to the master neuron is same for all values of $K_{sync}$ . In (f) it is seen that the slave neuron is almost synchronized with the master neuron at $K_{sync}=0.2$.  }
\label{syn}
\end{figure}

Here we have plotted the action potentials of both master and slave neurons with increasing Ksync value. We have shown the synchronization process here in the deterministic limit of A=20,000 and $I_{ext}=6.9$ $\mu A/cm^2 $. From figure \ref{syn} (a) we can see that the slave neuron is not at all in synchronization with the master neuron for Ksync=0.01. The master neuron response is same for the Ksycn values. Thus it is plotted only in the first and last value of the Ksync to compare it with slave response. As Ksync value increases the slave neuron starts firing initially at a random rate and then gradually the rate becomes equal to the master neuron at Ksync= 0.2. Although there exists little phase lag between the two neurons at Ksync= 0.2 which gradually eliminates at higher Ksync values. 
\vspace{0.35cm}

\subsection{Energetics of the unidirectional synchronization}
\begin{figure}[h!]
\centering
\includegraphics[width=9.0 cm,keepaspectratio]{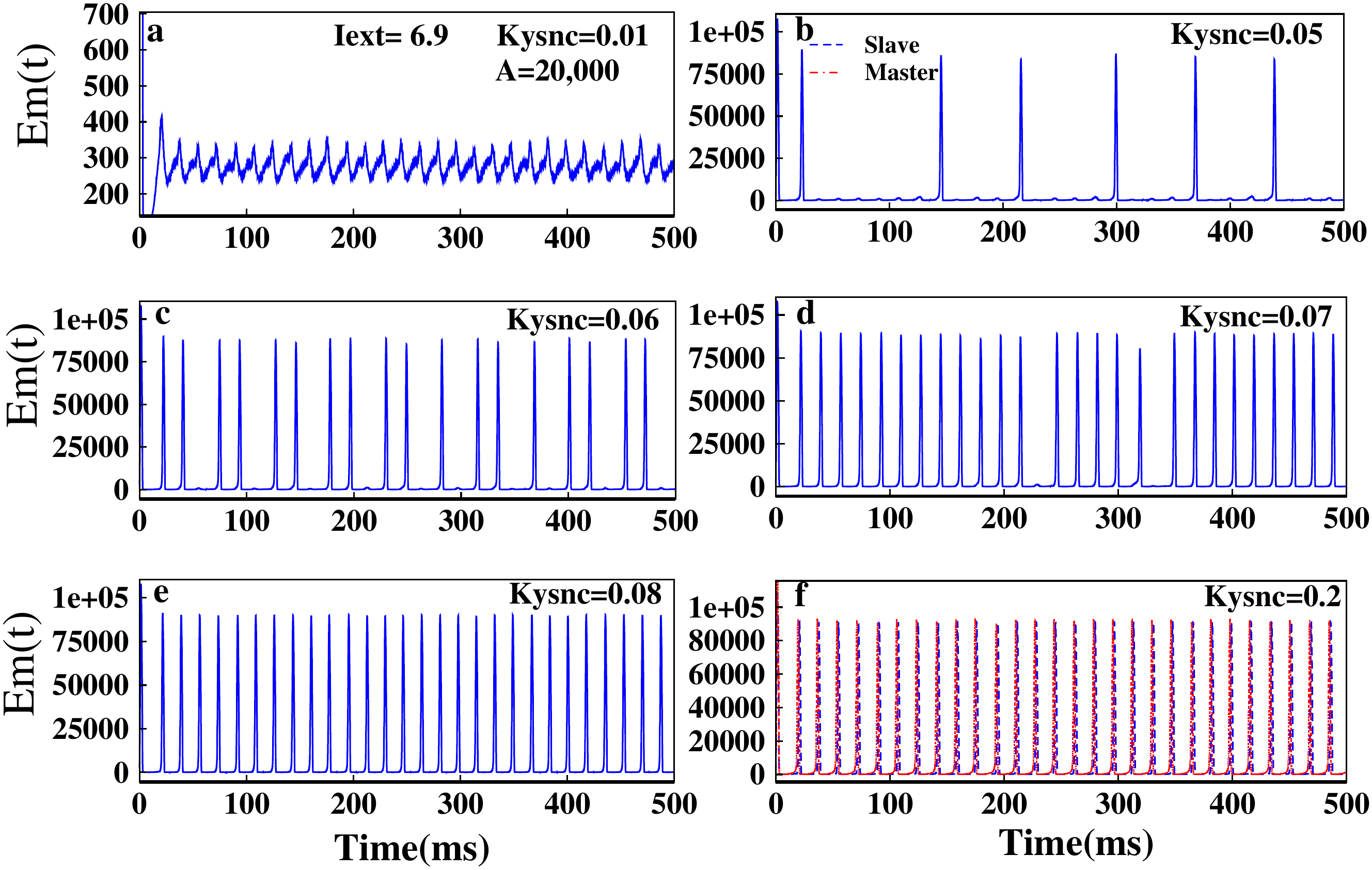}
\caption{Metabolic consumption of energy, Em(t) of slave neuron at different $K_{sync}$ values, for A=20,000$\mu m^2$, $I_{stimulus}(t)=I_{ext}(=6.9)+\xi(t)$. In (f) red dashed curve which corresponds to the master neuron is same for  all values of $K_{sync}$. In (f) it is seen that the slave neuron is almost synchronized with the master neuron at Ksync=0.2.}
\label{syn1}
\end{figure}
Next we have shown the total metabolic energy consumption, Em(t), by all the three channels together with time. It is seen from figure \ref{syn1}(a) that at very low value of Ksync the average metabolic energy consumption of the slave neuron is very less. As soon as  Ksync increases the spiking activity increases and as a result. Interesting fact here is that for each spike the maximum of the Em(t) is almost same for almost all the spikes and for all Kync values. At Ksync=0.2 the Em(t) for both master and slave neurons becomes almost equal as seen from figure \ref{syn1}(f) . 

\begin{figure}[h!]
\centering
\includegraphics[width=9.0cm,keepaspectratio]{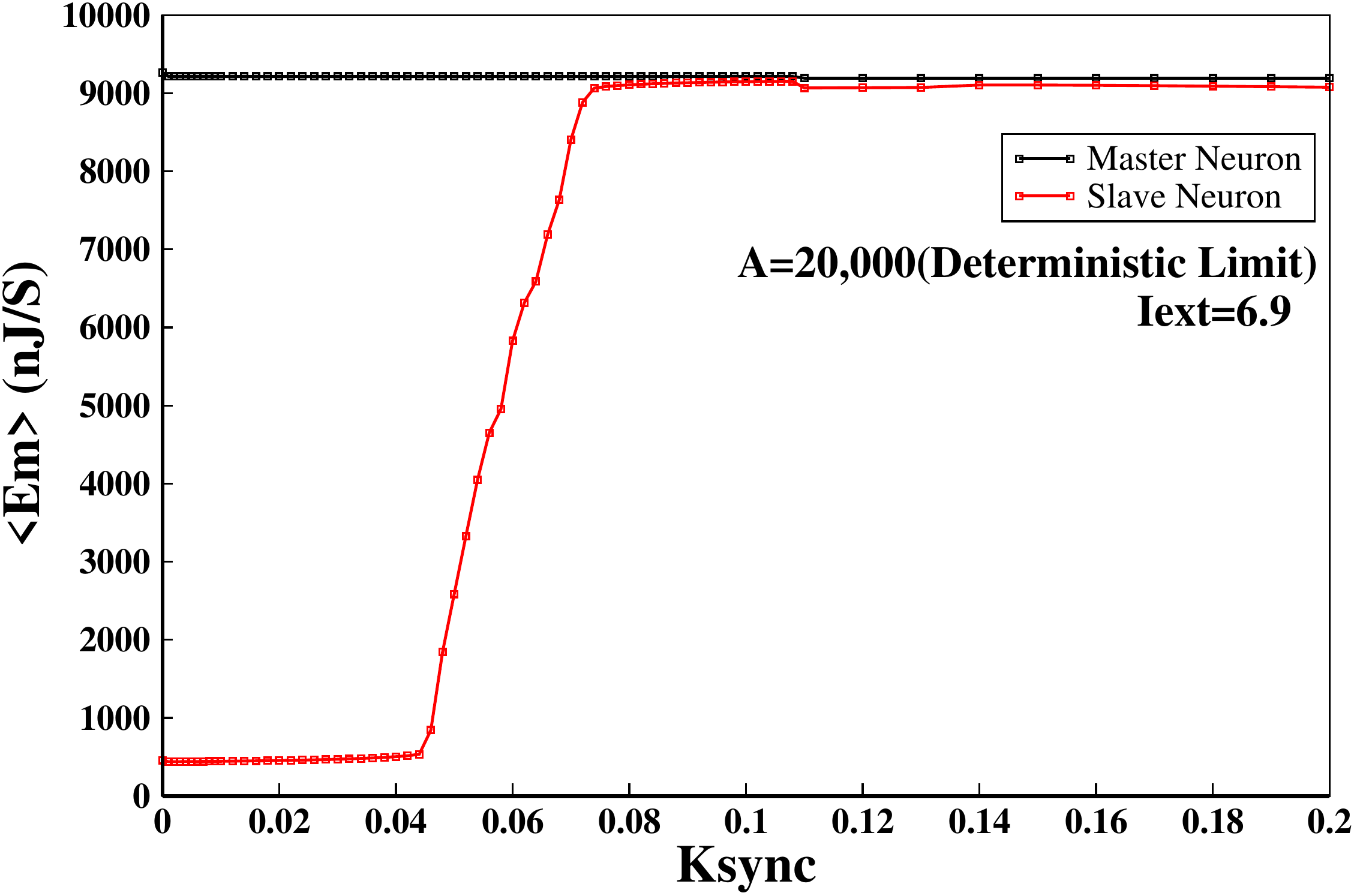}
\caption{The average metabolic energy consumption of the master and slave neuron is plotted here for different $K_{sync}$ values at A=20,000 $\mu m^2$ and $I_{stimulus}(t)=I_{ext}(=6.9)+\xi(t)$.  Each average was taken over very long time trajectories as shown in figure (\ref{syn1}). }
\label{patched}
\end{figure}

Next we have plotted the average consumption of metabolic energy with increasing Ksync value. As the spiking activity is stochastic in nature we take the average over a long period of time.  As it is unidirectional coupling the  coupling constant has no effect on master neuron. It is seen from figure \ref{patched} that the average total metabolic energy, $<Em>$, for the master neuron is around 9000 nJ/s. The average metabolic energy for slave neuron from around 350 nJ/s at $ K_{sync}= 0$ $ mS/cm^2$ gradually increases and eventually meets master neuron at $Ksync=0.1$.
\vspace{0.2cm}

\subsection{Effect of patch size variation on metabolic energy consumption}
Next we have studied the effect of patch size on the metabolic energy consumption. It is not possible for a neuron to have infinite numbers of ion channels in a finite patch size. As we have also seen that the patch size plays very important role. A low patch size  channel noise can alone give rise to spontaneous spiking activities. Thus studying the nature of metabolic energy consumption with variable patch size is very important.  
\begin{figure}[h]
\centering
\includegraphics[width=9.0 cm,keepaspectratio]{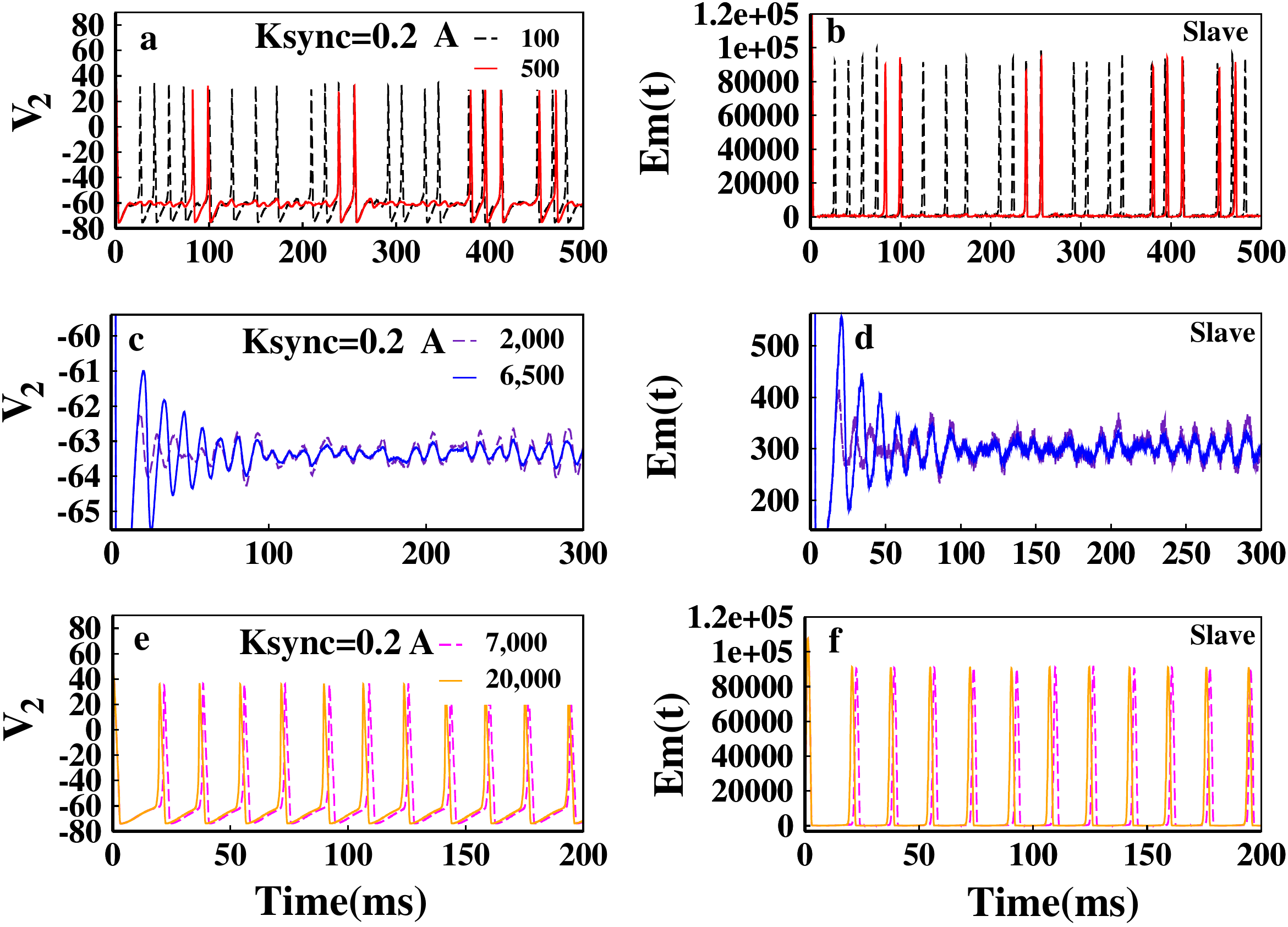}
\caption{Effect of patch size variation on action potentials and metabolic energy consumption is shown in this plot. Only the slave neuron responses have been plotted, as in $K_{sync}$ the master and slave neuron is almost synchronized. In the left column the slave action potentials and in the right corresponding Em(t)s are plotted.} 
\label{patch1}
\end{figure}
In the figure (\ref{patch1}) we have shown the action potential spikes and the metabolic energy consumption of slave neuron at Ksync=0.2 in left and right panels, respectively. In this Ksync value both the slave and master responds almost equally as seen from figure (\ref{syn}). Here we have shown the variation of patch size from A=100 to 20,000. In figure \ref{patch1}(a) one can see that with decreasing patch size channel noise plays very important role in action potential generation and consequently the metabolic energy consumption also increases as we decrease the patch size. Thus as we keep on increasing the patch size the spiking activity starts decreasing and suddenly and surprisingly at A=2000 the spiking activity totally vanishes and the corresponding metabolic consumption rate also falls down abruptly. This certain phenomenon continues with increasing value of A until  A=6500 is arrived as seen from figure \ref{patch1}(c) and (d). Again after A=7000 to  20,000 the spiking activity becomes almost similar to the deterministic result as seen from figure \ref{patch1}(e) and (f).

Thus there exist three different patch size ranges where the neurons behave differently. At very low patch size(e.g. A= 100-1500) the system dynamics and energetics are mainly governed by the channel noise or channel number fluctuations. In the mid range
(e.g. A =2000-6500) the neurons can not  even generate action potentials. This patch size range can be called as dead range. Then with increasing patch size(7000-20,000) the system gradually starts behaving as it behaves in deterministic limit. 

\begin{figure}[h]
\centering
\includegraphics[width=9 cm,keepaspectratio]{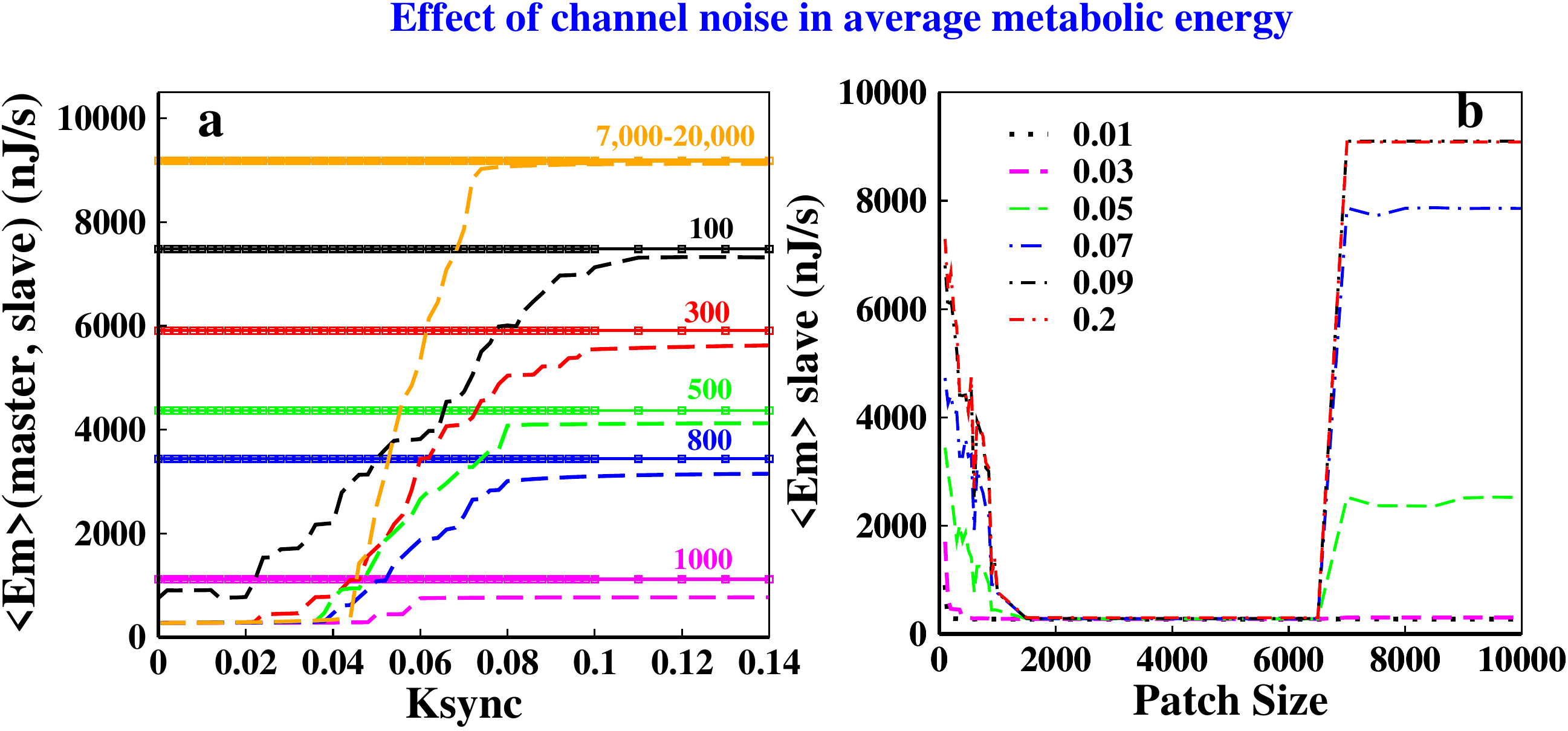}
\caption{In figure (a) the $<Em>$ for master and slave neurons are plotted with different $K_{sync}$ values for different patch size. In figure (b) $<Em>$ are plotted for different patch size for different $K_{sync}$ values.}
\label{patch2}
\end{figure}

Next we have done a detailed analysis of the effect of patch size on average metabolic energy, $<Em>$. In figure \ref{patch2}(a) we have ploted $<Em>$ with different Ksync values for different patch sizes. It is seen that with increasing patch size the average metabolic energy consumption for both master and slave neuron decreases. It continues to occur until around A=1500. After that for A=2000 to 6500 the $<Em>$ becomes very low and again after A=7000 to 20,000 $<Em>$ almost remains close to the deterministic average.  In figure \ref{patch2}(b) $<Em>$ is plotted with patch size for different Ksync values. The dependence on the patch size as depicted earlier is now evident from this picture. There exist  three distinct dependency of patch size on average metabolic energy consumption. The very low and very high patch size region shows opposite dependence of patch size and the mid range acts as an dead or inactive zone. For an unidirectionally coupled neuron system this mid range of patch size for both master and slave neurons probably are not a good combination for the generation of action potential unidirectionally. 

From figure \ref{patch2}(a) it is also seen that with decreasing patch size, the Ksync value at  which the $<Em>$ of master and slave neuron matches, is gradually right shifted. This means the synapse is now finding difficulties in bringing the two neuron in synchronization. Thus with decreasing patch size the synapse needs to work more to synchronize the energies of the two neurons. 
\vspace{0.5cm}

\section{Effect of Ion Channel Blocking Drugs and the Validation of the Result}
\label{s5}

Here we have studied the effect of three different types of drug blockers on the spiking activity and metabolic energy consumption rate. We have taken examples of three types of drug blocking. Among them one is sodium only blockers, e.g. tetrodotoxin (TTX), second one is potassium blockers, e.g. tetraethylammonium (TEA) and the third type is total blockers\cite{hille, kp, kp1}. These drugs selectively blocks either sodium or potassium channels and thus the available number of channels left for ion conduction gets reduced. This is why  drug binding study here actually falls under the study of patch size effect.  Here we have considered that both the master and slave neuron have equal number of ion channels left after the addition of drugs or we can say that both of the neurons are similarly affected by the drugs\cite{hangii4}. For considering the effect of drugs the sodium and potassium conductance modifies as follows\cite{hangii4},
\begin{equation}
G_K(t)= g^{max}_K x_K n^4\hspace{1cm} \text{and} \hspace{1cm}G_{Na}(t)= g^{max}_{Na} x_{Na} m^3h,
\label{drug}
\end{equation}
where $x_K$ and $x_{Na}$ are the fractions of working ion channels which are  not blocked by drugs among the overall number of potassium channels, $N_K$ , or sodium ion channels, $N_Na$ ,  respectively.
Thus equations (\ref{hh}), (\ref{lang1}), (\ref{lang2}) and (\ref{drug}) form a stochastic Hodgkin-Huxley model takes drug blocking into account. This set of equations with very high patch sizes, e.g. A=20,000 corresponds to the deterministic limit.

\subsection{Sodium channel blockers}

\begin{figure}[h!]
\centering
\includegraphics[width=9.0 cm,keepaspectratio]{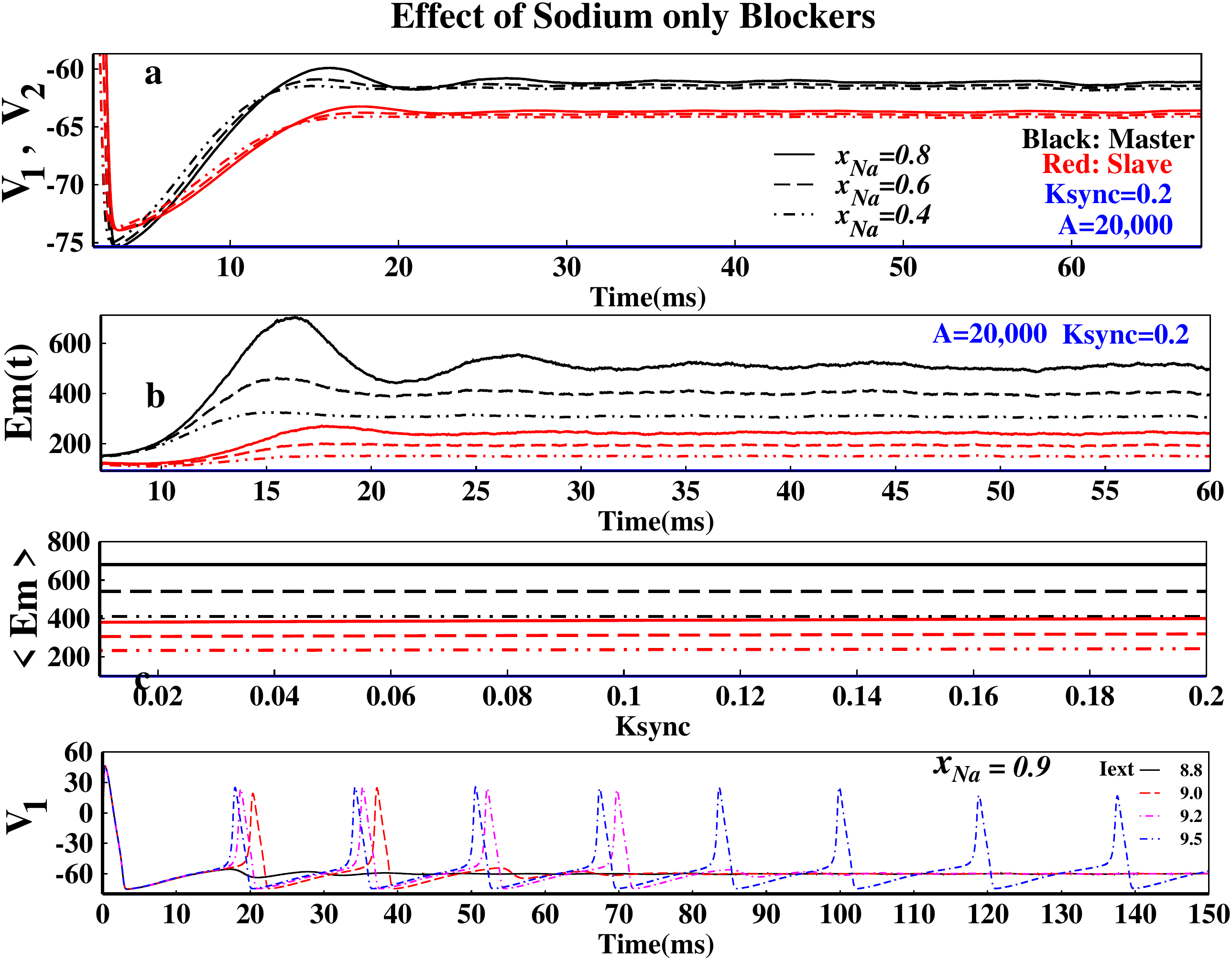}
\caption{Effect of Sodium Blockers on action potential and metabolic energy consumption, deterministic limit. In figure (a) the master and slave neurons for have been plotted for $x_{Na}=$ 0.8, 0.6 and 0.4. In (b) corresponding $Em(t)$ and in figure (c) $<Em>$ is plotted for different $K_{sync}$ values. In  figure (d) the master neuron's action potentials are shown for $I_{ext}=8.8$,  9.0, 9.2 and 9.5 for $x_{Na}=0.9$.   } 
\label{sod}
\end{figure}

Here we have studied the effect of sodium blockers on action potential and metabolic energy consumption. We have found very drastic effect of sodium blocks here. Here it is shown that even a very minute change in number of available sodium channel the repetitive spiking action totally vanishes surprisingly. We have studied the effect of sodium blockers for $x_{Na}=0.8, 0.6, $ and $0.4$ as seen from figure \ref{sod}(a). The master and slave action potentials have been shown here. We can see that at Ksync=0.2 there is no considerable change in the trends of action potentials for different $x_{Na}$ values but importantly the synchronization between mastr and slave is obviously destroyed. 
The corresponding Em(t) is shown in figure \ref{sod}(b) and the $<Em>$ with Ksync has been plotted in  figure \ref{sod}(c). From these two figures we can see that the average metabolic consumption rate drastically falls. Thus minute variation in sodium ion channels in a particular external current will cause drastic effect on action potential train. However we have shown that with these fractions of sodium channel the spiking activities can be seen if the $I_{ext}$ is increased gradually as seen from figure \ref{sod}(d).  Here the $I_{ext}$ used are 8.8, 9.0, 9.2, 9.5 for $x_{Na}=0.9$. Here also we have seen that the master and slave neurons are not synchronized. Also as we decrease the fraction $x_{Na}$, such as  $x_{Na}=0.8$, higher magnitude of $I_{ext}$ is required to resurrect the spikes again. 

\subsection{Potassium channel blockers} 
\begin{figure}[h!]
\centering
\includegraphics[width=9.0 cm,keepaspectratio]{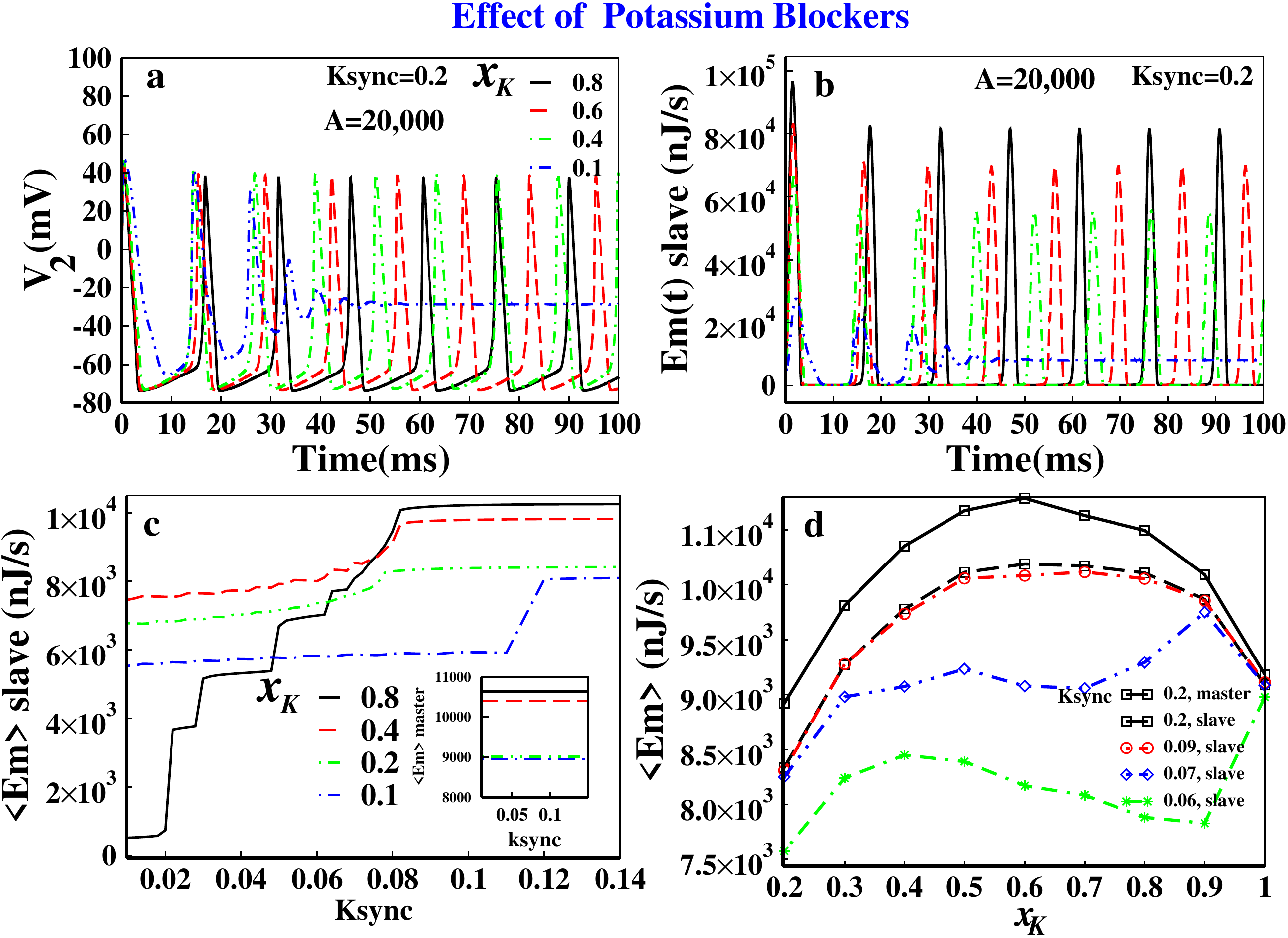}
\caption{Effect of potassium blockers on metabolic energy consumption in the deterministic limit. In figure (a), (b) and (c) the slave neuron action potentials, Em(t) and $<Em>$ is plotted for $x_{K}$=0.8, 0.6, 0.4 and 0.1, respectively. In figure (d) $<Em>$ is plotted with $x_K$ for different $K_{sync}$ values. } 
\label{pot}
\end{figure}

Next we have studied the effect of potassium blockers for $x_{K}$ = 0.8, 0.6, 0.4 and 0.1.  Unlike sodium blockers the potassium blockers have no instant drastic effect on action potential of both master and slave neurons. In figure \ref{pot} the effect of potassium blockers have been shown for Ksync=0.2 where both master and slave neurons remain almost synchronized. Thus in figure \ref{pot} we have shown only the effect on slave neuron.
\vspace{0.2cm}

From figure \ref{pot}(a) it is seen that initially with decreasing $x_K$ value from $x_K=1.0$ the spiking activity increases and continues to increase until a certain $x_K$ and then gradually decreases. Here we have shown the picture for first 100 ms only to avoid congestion. A similar trend is observed in the metabolic energy consumption rate as seen from figure \ref{pot}(b).  With increasing drug concentration or decreasing available potassium channels the maximum of spikes decreases as seen from figure \ref{pot}(b). In figure \ref{pot} (c) we find that at $x_K=$ 0.8 and 0.4 the $<Em>$ at Ksync=0.2 are almost similar. It is also seen that at very low Ksync such as 0.01 or 0.02 etc the $<Em>$ of $x_{K}=$0.4, 0.2 or 0.1 of the slave neuron starts from relatively much higher magnitude than $x_K=$ 0.8 or 1.0. For better understanding we have plotted the $<Em>$ for a particular Ksync value for different $x_K$ in figure \ref{pot}(d). Here we can clearly see for Ksync=0.2, the average metabolic consumption initially increase with decreasing $x_K$ until $x_K=0.6$ is reached. After that $<Em>$ again decreases both for master and slave neurons. This initial increase in spiking activity is attributed to the noise enhanced spiking activity(as reported earlier\cite{hangii4}). Figure \ref{pot} (d) shows that the noise enhanced spiking activity is also evident  in metabolic energy calculations. This is also a validation of the simulation result.  Now as we decrease the Ksync value we see that the noise enhanced phenomena shows different nature in different Ksync values.

\subsection{Total blockers} 
\begin{figure}[h]
\centering
\includegraphics[width=9.0 cm,keepaspectratio]{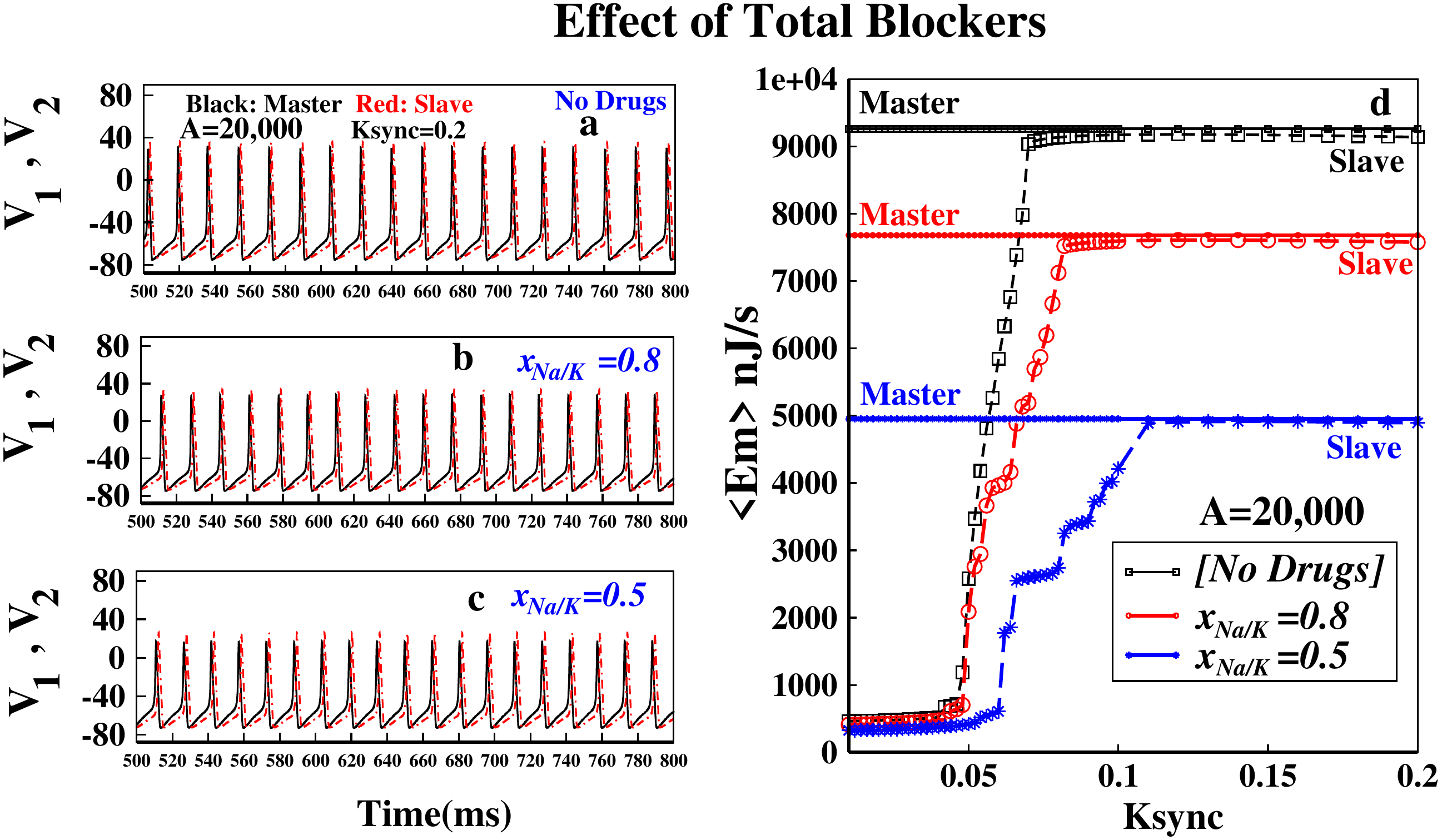}
\caption{Effect of Total blockers on average metabolic energy consumption in the deterministic limit. In the left panel from (a)-(c) the effect of total blockers on action potentials for both the master and slave neurons are plotted but not that much observable effect or difference is seen. On the contrary in figure (d) the blockers have pretty distinguishable effect on energetic. } 
\label{tot}
\end{figure}

Here we study the effects of drugs that may block both sodium and potassium channels. For simplicity we keep both  $x_{Na}=x_K=x_{Na/K}$= 0.8 and 0.5  in figure \ref{tot} with ksync=0.2. In the left panel the transient action potentials have been shown in presence of no drug and for $x_{Na/K}$= 0.8 and $x_{Na/K}$= 0.5 in figure \ref{tot} (a), (b), (c) respectively. It is interesting to see that there are not much kinetic differences between the three different concentrations of drug except phase difference and may be slight difference in spiking rate. But for energetic analysis it is seen that all the three types are energetically very distinct process as seen from figure \ref{tot}(d). It is seen that with increase in the drug concentration or decrease in the available number of ion channels available the average metabolic energy consumption decreases. Also one thing can be noticed that with increasing number of channel blockers the value of Ksync at which both master and slave neuron's average energy consumption becomes equal, is gradually right shifted. This means  synapse also finds it difficult to bring to neurons in synchronization. Synapse needs to work more with increasing number of blocked ion channels.

\section{Conclusion}
\label{s6}

In this work we have studied the energetic cost of the signaling activity through  generation of spikes for single neuron as well as  unidirectionally  coupled neurons with an electrical synapse.  The basic ingredient of internal noise arising from individual stochastic dynamics of the ion channels are utilized here to characterize  spiking  for a critical patch size.
However, here our main focus is on the effect of the patch size on metabolic energy consumption of the master and slave neurons during the process of synchronization. The detailed results can be summarized as follows.
\bigskip

(1) To standardize the existing system parameters of squid giant axon we have shown that for a single neuron without the presence of any external current the deterministic limit is reached within very low patch size(A=150 $\mu$m$^2$), but for  two coupled neurons with an external current( such as  6.9 $\mu A/cm^2 $) a very large number of system size( e.g. A=20,000 and A=5,000) for a critical value of external current($I_{ext}= $10.0$\mu A/cm^2$) is required to produce deterministic result. With increasing  external current the required patch area decreases to reach the deterministic limit.
\bigskip

(2) There exists three different ranges of patch size where the coupled system behaves in a very different manner.  At very low patch size range (e.g. A= 100-1500) the system dynamics and energetics are mainly governed by the channel noise or channel number fluctuations. In the mid range (e.g. A =2000-6500) the system fails to  respond properly. It can not generate trains of action potentials in this range which 
adequately can be  called as dead range. Then with increasing patch size the system gradually starts behaving as it should follow in deterministic limit.
\bigskip

(3) The range of patch size   where channel noise predominates, with increasing patch size metabolic energy consumption decreases as channel noise decreases and with increase in patch size the  spontaneous spiking activity decreases. Next in the mid range  as there exists no spiking trains, the metabolic consumption of energy falls drastically. Then again in high patch size range with increasing spiking activity both master and slave neuron starts firing pattern close to deterministic limit with increasing metabolic energy consumption.
\bigskip

(4) We have found very interesting effect of sodium, potassium and total blockers on both spiking activity and energetic costs. Both the synchronization process and the spiking activity of action potentials are greatly affected by sodium channel blockers. A minute change in the number of available open sodium channels totally destroys the spiking activity.  There is no considerable change in action potential  with increasing number of blocked sodium channels as well, but there exists minute changes visible in metabolic consumption. Although with increasing external current spikes resurrects but neurons remain synchronized.
\bigskip

(5) There exists a strikingly different result for potassium channel blockers. Initially with increasing concentration of drug the metabolic energy increases and reaches a maximum for a particular $K_{sync}$ value, spiking activity also increases, which is attributed to the noise enhanced spiking phenomena. Then again with decreasing  number of available potassium channel metabolic consumption decreases. This noise enhanced spiking phenomenon has been validated with the simulated data. 
\bigskip

(6) For  total blockers  with decreasing number of available ion channels the average metabolic energy consumption decreases which can be understandable from the property of internal noise dynamics. Comparing all types of blockers we have found that with increasing number of blocked ion channels the synaptic efficiency plays a vital role to bring synchronization between two neurons.
Generally speaking the synapse  finds it difficult to bring the neurons in synchronization in presence of drug or in other words  the synapse needs to be more conducting with increasing number of blocked ion channels.

\end{document}